\documentclass[twocolumn,showpacs,preprintnumbers,amsmath,amssymb,prl,superscriptaddress]{revtex4}

\usepackage{amsmath,amssymb,amsthm,color}
\usepackage{graphicx}
\usepackage{dcolumn}

\newcommand {\beq}{\begin{eqnarray}}
\newcommand {\eeq}{\end{eqnarray}}

\begin{document}

\preprint{CALT-68-2837, IPMU11-0094}

\title{Comments on Worldsheet Description of the Omega Background}

\author{Yu Nakayama}

\affiliation{California Institute of Technology, 452-48, Pasadena, California 91125, USA}

\author{Hirosi Ooguri}

\affiliation{California Institute of Technology, 452-48, Pasadena, California 91125, USA}
\affiliation{Institute for the Physics and Mathematics of the Universe,  \\ Todai Institutes for Advanced Study,
University of Tokyo, \\ 
5-1-5 Kashiwanoha, Kashiwa, Chiba 277-8583, Japan}


\begin{abstract}
Nekrasov's partition function is defined on a flat bundle of $\mathbb{R}^4$ 
over $\mathbb{S}^1$ called the Omega background. 
When the fibration is self-dual, the partition function is 
known to be equal to the topological string partition function, 
which computes scattering amplitudes of self-dual gravitons and graviphotons 
in type II superstring compactified on a Calabi-Yau manifold. We propose a 
generalization of this correspondence when the fibration is not necessarily 
self-dual. 

\end{abstract}

\maketitle

\section{1. Introduction}

The topological string theory has been used to compute a variety of 
physical observables in type II superstring theory and M theory 
compactified on a Calabi-Yau manifold. The genus-$g$ contribution 
$F_g$ to the partition function is related to the scattering amplitude 
of $2$ self-dual gravitons and $(2g-2)$ self-dual graviphotons in the zero 
momentum limit and computes the corresponding F-term in the low energy effective 
action in four dimensions \cite{Bershadsky:1993cx,Antoniadis:1993ze}. 
The all genus partition function can be used to count 
the number of BPS particles in M theory compactified to 
five dimensions \cite{Gopakumar:1998ii}. Its absolute value squared is related to the 
generating function for BPS states in 
type II string theory compactified to four dimensions \cite{Ooguri:2004zv}. 

For a five-dimensional rigid ${\cal N}=2$ supersymmetric theory on $\mathbb{R}^4 \times \mathbb{S}^1$, 
which can be realized, for example, by M theory on a non-compact 
Calabi-Yau 3-fold $X$, Nekrasov's partition function is defined by 
\begin{equation}\label{nekrasov}
Z(\epsilon_+, \epsilon_-) 
= {\rm Tr}\ (-1)^F e^{ -2\epsilon_- J_-^3}e^{ -2\epsilon_+(J_+^3 + J_R^3)}e^{-\beta H}, 
\end{equation}
where the trace is taken over the Hilbert space of the theory on $\mathbb{R}^4$, 
$J_\pm^3$ are the Cartan generators of $SU(2)_\pm$ subgroups of the $SO(4)$ 
rotation of $\mathbb{R}^4$, $J_R^3$ is the Cartan generator of the $SU(2)$ R symmetry,
and $H$ is the translation generator for $\mathbb{S}^1$ whose radius is set by $\beta$ 
\cite{Nekrasov:2002qd,Losev:2003py,Nekrasov:2003rj}. It can be regarded as a vacuum amplitude
on a Melvin-type geometry, where the $\mathbb{R}^4$ is fibered over the $\mathbb{S}^1$
(with additional twist by the R symmetry). This geometry is called the
Omega background.

When $\epsilon_+=0$ and when the ${\cal N}=2$ theory arises from 
M theory on $\mathbb{R}^4 \times \mathbb{S}^1 \times X$ for some Calabi-Yau 3-fold $X$, 
this BPS state counting problem is related to the topological string partition function as, 
\begin{equation}
\label{unrefined}
Z(\epsilon_+=0, \epsilon_- = g_s) =
 \exp\left( \sum_{g = 0}^\infty g_s^{2g-2} F_{g}(X) \right),
\end{equation}
where $g_s$ is the genus counting parameter, and $F_g(X)$ is the genus-$g$ topological string partition function for $X$ 
\cite{Gopakumar:1998ii}. When $\epsilon_+ \neq 0$, 
Nekrasov's partition function can still be expanded as,
\begin{equation}
\label{toppartition}
Z(\epsilon_+, \epsilon_-) =
 \exp\left( 
  \sum_{g=0}^\infty g_s^{2g-2} F_g(b) \right), 
\end{equation}
where $\epsilon_+ =\frac{i}{2}g_s (b+b^{-1})$, $\epsilon_- = \frac{i}{2} g_s (b-b^{-1})$. 

It is natural to ask if there is a worldsheet description of 
the coefficients $F_{g}(b)$.  
For toric Calabi-Yau manifolds, generalizations of the topological vertex for $F_{g}(b)$ 
have been proposed \cite{Awata:2005fa,Iqbal:2007ii}. 
For Calabi-Yau manifolds realized by fibrations over Riemann surfaces, 
deformed matrix models have been proposed \cite{Dijkgraaf:2009pc}.

Recently, Antoniadis, et al. proposed a worldsheet definition of $F_{g}(b)$ 
\cite{Antoniadis:2010iq}. It involves scattering amplitude $A_{n_G, n_F}$ among $2$ gravitons, $(2n_G-2)$ graviphotons, and $2n_F$  
gauge fields in vector multiplets.
 When there is a heterotic string dual on 
$T^2 \times K3$, it was proposed that the $2n_F$ gauge fields belong to 
the linear dilaton multiplet $S$, which is a vector multiplet in the heterotic
string theory. The scattering amplitude in the heterotic dual has been
computed in \cite{Morales:1996bp} in the limit of $S \rightarrow \infty$. 
It is related to certain higher derivative terms in $\mathcal{N}=2$ supersymmetric
effective action whose superconformal structure was studied in \cite{deWit:2010za}.
In this paper, we propose a different interpretation of the 
Omega background. 

As we will review in the next section, even when $\epsilon_+=0$, 
the Omega background for Nekrasov's partition function (\ref{nekrasov})
and the self-dual graviphoton background used to compute the topological string
partition function $F_g$ are similar but not identical. In this paper, 
we will generalize each of these backgrounds for $\epsilon_+ \neq 0$. 

This paper is organized as follows. In section 2, we review the geometry of
the Omega background and explain why Nekrasov's partition function is equal to
the topological string partition function when $\epsilon_+ = 0$. In
section 3, we study configurations with non-zero field strengths
for the graviphoton and vector multiplet gauge fields and examine the amounts of
supersymmetry they preserve. In section 4, we
discuss their realizations in the heterotic string on $T^2 \times K3$ and propose
string theory configurations corresponding to the Omega background
and the topological string computation when $\epsilon_+ \neq 0$. 
In section 5, we discuss interpretation of these backgrounds in the context of 
type II superstring theory. 

\section{2. Review of the Omega background}

\subsection{2.1 Melvin-type geometry}

The Omega background is a Melvin-type geometry with the metric,
\begin{equation}
\label{Melvin}
ds^2 = (dx^\mu + \Omega^\mu d\theta)^2 + d\theta^2, 
\end{equation}
where $x^\mu$ and $\theta$ are coordinates on $\mathbb{R}^4$ and $\mathbb{S}^1$ respectively,
and $\Omega_\mu$ is characterized by the equation, 
\begin{equation}
\label{fieldstrength}
 \mathcal{F} = d\Omega = \epsilon_1 dx^1 \wedge dx^2 + \epsilon_2 dx^3 \wedge dx^4. 
\end{equation}
The parameters $\epsilon_\pm$ in Nekrasov's partition function (\ref{nekrasov}) 
are defined by,
\begin{equation}
  \epsilon_\pm = \frac{1}{2}(\epsilon_1 \pm \epsilon_2). 
\end{equation}
A more invariant way to define $\epsilon_\pm$ is to express the field strength in 
the spinor notation $(\mathcal{F}_{\alpha\beta}, \mathcal{F}_{\dot \alpha \dot \beta})$ ($\alpha,\beta$ = 1,2;
$\dot \alpha, \dot \beta = 1,2$) and write
\begin{equation}
 \epsilon_-^2 = -{\rm det} \mathcal{F}_{\alpha\beta},~~
\epsilon_+^2 = -{\rm det} \mathcal{F}_{\dot \alpha \dot \beta}. 
\end{equation}
Namely, $\epsilon_-$ and $\epsilon_+$ are related to self-dual and anti-self-dual components
of the field strength. 

When $\epsilon_+ \neq 0$, we need to perform the R twist simultaneously
in order to preserve some fraction of supersymmetry. 
This requires, in particular, that the internal Calabi-Yau 3-fold be non-compact
and the Newton constant in four dimensions is zero, resulting in a rigid ${\cal N}=2$ theory. 

The Melvin-type geometry (\ref{Melvin}) is locally flat and should be contrasted with the 
Kaluza-Klein ansatz, 
\begin{equation}
\label{KK}
 ds^2 = dx^2 + (d\theta + A_\mu dx^\mu)^2. 
\end{equation}
If we choose the Kaluza-Klein gauge field
$A_\mu$ and the corresponding $\Omega_\mu$ for the Melvin-type geometry so that
\begin{equation}
  A_{\alpha\dot \alpha} = \Omega_{\alpha\dot\alpha}= F_{\alpha\beta}\ x^\beta_{~\dot \alpha}
    +F_{\dot \alpha \dot \beta}\ x_\alpha^{~\dot\beta},
\end{equation}
the two metrics (\ref{Melvin}) and (\ref{KK}) become equal at the linear order in $x$.
However, they start to differ at $\mathcal{O}(x^2)$ \cite{Losev:2003py}. 
In the self-dual case $\epsilon_+=0$, the relation between 
computations in the two backgrounds 
was pointed out in \cite{Gopakumar:1998ii}.
In this paper, we propose a generalization of this relation
 for non-self-dual field strengths. 

We can also start with a $(1,0)$ theory in six dimensions (for example, the
heterotic string theory on $K3$) and compactify it on a two-torus $T^2$ with
the metric, 
\begin{equation}\label{heteromelvin}
ds^2 = (dx^\mu + \Omega^\mu dz + \bar{\Omega}^\mu d\bar z)^2 + dz d\bar z,
\end{equation}
where $(z,\bar z)$ are coordinates on $T^2$. In this case, $\epsilon_\pm$ given
by $d\Omega$ as in (\ref{fieldstrength}) are complexified. It turns out 
that, by supersymmetry, Nekrasov's partition function depends holomorphically on these parameters 
\cite{Nekrasov:2003rj,Ito:2010vx}. The corresponding Kaluza-Klein ansatz is
\begin{equation} \label{consthetero}
 ds^2 = dx^2 + (dz + 2\bar{A}_\mu dx^\mu) (d\bar z + 2A_\mu dx^\mu). 
\end{equation}

To preserve supersymmetry, the partition function (\ref{nekrasov}) is
twisted by the R symmetry. Correspondingly, we should
turn on the flat gauge field coupled to the $SU(2)$ R charge,
\begin{equation} \label{twist} 
A^{\ i}_j = \mathcal{A}^{\ i}_j dz + \bar{\mathcal{A}}^{\ i}_j d\bar{z} ,
\end{equation}
where $i = 1,2$ is the $SU(2)$ R symmetry index and
$\mathcal{A}^{\ i}_j$ has eigenvalues $\pm \epsilon_+$. 
In the heterotic string on $\mathbb{R}^4 \times T^2 \times K3$, 
the R symmetry is the isometry of $K3$ along the Reeb vector.
In this case, $A^{\ i}_j$ is given by the Kaluza-Klein 
gauge field associated to this isometry.

In the heterotic string theory on $T^2 \times K3$, there are three universal vector multiplets:
$S$ (dilaton), $T$ (complex structure moduli of $T^2$), and $U$ (K\"ahler moduli of $T^2$). 
Combining them with the graviphoton $G$, there are four gauge fields, $A^G, A^S, A^T, A^U$, which correspond to 
the off-diagonal elements of the metric $(g_{\mu z}, g_{\mu \bar z})$ and the 
NS-NS anti-symmetric tensor $(B_{\mu z}, B_{\mu \bar z})$, where $\mu =1,2,3,4$
are for $\mathbb{R}^4$ and $(z, \bar z)$ are coordinates on $T^2$. 
By inspecting their vertex operators, their precise relations are found to be,
\begin{equation}\label{GSTU}
\begin{split}
& F^G_- = (\partial_{[\mu,} g_{\nu] z} + \partial_{[\mu,} B_{\nu] z})_-
,~~ F^{\bar{G}}_+ = (\partial_{[\mu,} g_{\nu] \bar z} + \partial_{[\mu,} B_{\nu] \bar z})_+, \\
& F^T_- = (\partial_{[\mu,} g_{\nu] z} - \partial_{[\mu,} B_{\nu]  z})_-
,~~ F^{\bar{T}}_+ = (\partial_{[\mu,} g_{\nu] \bar z} - \partial_{[\mu,} B_{\nu] \bar z})_+, \\
& F^U_- = (\partial_{[\mu,} g_{\nu] \bar z} - \partial_{[\mu,} B_{\nu] \bar z})_-
,~~ F^{\bar{U}}_+ = (\partial_{[\mu,} g_{\nu] z} - \partial_{[\mu,} B_{\nu] z})_+, \\
& F^S_- = (\partial_{[\mu,} g_{\nu] \bar z} + \partial_{[\mu,} B_{\nu] \bar z})_-
,~~ F^{\bar{S}}_+ = (\partial_{[\mu,} g_{\nu] z} + \partial_{[\mu,} B_{\nu] z})_+, \\
\end{split}
\end{equation}
where $F_-$ and $F_+$ refer to self-dual and anti-self-dual components of the gauge field strength. 
We see, for example, that the self-dual component of $(\partial_{[\mu,} g_{\nu] \bar z} - \partial_{[\mu,} B_{\nu] \bar z})$
is $F^T_-$, but its anti-self-dual component is $F^{\bar{U}}_+$. 

If we set $\bar{A}_\mu=0$ in (\ref{consthetero}), anticipating that 
Nekrasov's partition function depends holomorphically on $\epsilon_\pm$,
\begin{equation}
 ds^2 = dx^2 + dz (d\bar z + 2A_\mu dx^\mu) = dx^2  + 2 A_\mu dx^\mu dz + dz d\bar z. 
\end{equation}
The field strength components for $A_\mu$ are identified with those for 
the graviphoton and the $STU$ gauge fields as, 
\begin{equation}\begin{split}
F_{\alpha\beta} = F_{\alpha\beta}^G + F_{\alpha\beta}^T , \cr 
F_{\dot \alpha \dot \beta} = F_{\dot \alpha\dot \beta}^{\bar{U}} + F_{\dot \alpha \dot \beta}^{\bar{S}}.
\end{split}\end{equation}
Moreover, if we require that the NS-NS $B$ field is zero, 
we have $F^G = F^T$ and $F^{\bar{U}} = F^{\bar{S}}$, and thus,
\begin{equation}\label{GSTUidentification}
\begin{split}
F_{\alpha\beta}^G = F_{\alpha\beta}^T = \frac{1}{2} F_{\alpha\beta}, \cr 
F_{\dot \alpha\dot \beta}^{\bar{U}} = F_{\dot \alpha \dot \beta}^{\bar{S}}
= \frac{1}{2} F_{\dot \alpha \dot \beta}.
\end{split}
\end{equation}
To the leading order in $\epsilon_\pm$, we can identify this as the Melvin-type geometry. 
Corrections in higher order in $\epsilon_\pm$ will be discussed in section 3. 
For the Omega background, we need to perform the R twist to preserve fraction of supersymmetry. 
As we will discuss in sections 3 and 4, it is realized by
turning on the Fayet-Iliopoulos (FI) terms in the corresponding supergravity backgrounds.

In the self-dual case ($\epsilon_+^2= -{\rm det} F_{\dot \alpha \dot \beta} = 0$), the one-loop partition function in the Melvin background 
\eqref{heteromelvin} was shown to reproduce the all genus topological string amplitude in the limit of $S \rightarrow \infty$.
Let us discuss this in more detail now.

\subsection{2.2 Relation to the topological string}

The genus-$g$ contribution to the topological string partition function on a Calabi-Yau manifold $X$
is related to a scattering amplitude $A_g$ with $2$ gravitons and $(2g-2)$ graviphotons in type II
superstring theory compactified on $X$, all
 at zero momentum and with a particular self-dual polarization 
 \cite{Bershadsky:1993cx,Antoniadis:1993ze}, as
\begin{equation} 
 A_g = (g!)^2 F_g. 
 \end{equation}
The combinatorial factor $(g!)^2$ comes from the superspace integral to relate the F-terms computed by 
$F_g$ to the low energy effective action of type II superstring theory together with the particular choice of the polarization. 
 
In the heterotic string theory on $T^2 \times K3$, the dilaton $S$ belongs to a vector multiplet. 
When type II theory on $X$ has a heterotic string dual, 
and in the limit $S \rightarrow \infty$, the topological string partition function
can be computed by the heterotic string one-loop as \cite{Antoniadis:1995zn}
\begin{equation}\label{hetoneloop}
\begin{split}
\sum_{g=0}^\infty  \epsilon_-^{2g-2} F_g  & 
=\sum_{g=0}^\infty \frac{1}{(g!)^2}\epsilon_-^{2g-2} A_g \\
&=\frac{1}{\epsilon_-^2}\int d\tau^2 \langle e^{-\Delta S_{\mathrm{eff}} } \rangle_{R^4 \times T^2} \times Z_{K3} 
 \end{split}
 \end{equation}
with 
\begin{equation}\label{hetac}
\begin{split}
\Delta S_{\mathrm{eff}} =  
&\epsilon_- \int \left[ ( X^1 \partial Z - \chi^1 \psi_Z) \bar{\partial} X^2 \right. \\ 
&~~~~+ \left.  (\bar{X}^1 \partial Z - \bar{\chi}^1 \psi_Z) \bar{\partial} \bar{X}^2\right] \ .
\end{split}
\end{equation}
In (\ref{hetoneloop}),  $\tau$ is the moduli of the worldsheet torus and the integral is over the fundamental domain.
In (\ref{hetac}), the integral is over the worldsheet torus, 
$X^1, X^2$ are holomorphic coordinates on $\mathbb{R}^4$ (with their fermionic partner $\chi^1, \chi^2$), $Z$ is a holomorphic coordinate on $T^2$ (with its fermionic partner $\Psi_Z$), and $Z_{K3}$ represents contributions
from the $K3$ part of the target space. We note that the effective action \eqref{hetac} is invariant under the target space supersymmetry we will discuss in section 3.1.1.

In the heterotic string theory, elementary BPS particles are in ground states in the
left-moving (supersymmetric) sector and in the tower of bosonic oscillator excitations 
in the right-moving (bosonic) sector. From (\ref{hetoneloop}), one can deduce that each 
of such BPS multiplets contributes to the right-hand side of
(\ref{hetoneloop}) as,
\begin{equation}\label{Schwinger}
 \sum_{g=0}^\infty \epsilon_-^{2g-2} F_g 
  = (-1)^{2j} \int_0^\infty
  \frac{dt}{t} \frac{ {\rm tr}\ e^{-4t \epsilon_- J_-^3}}{(\sinh \epsilon_- t)^2} e^{-t \mu}, 
  \end{equation}
where $j$ is the total spin of the BPS particle and $\mu$ is its central charge for 
the ${\cal N}=2$ supersymmetry \cite{Gopakumar:1998ii}. 
Here, the trace is taken only over the $SU(2)_-$ representation of 
the one-particle BPS states.
An important observation was that 
the right hand side of \eqref{Schwinger} is exactly the contribution of
each of the BPS multiplets to ${\rm Tr} (-1)^{2j} e^{-2\epsilon J_-^3} e^{-\beta H}$ 
from the one-loop computation  \cite{Gopakumar:1998ii,Nekrasov:2002qd}.
This was the first evidence for the identification,
\begin{equation}\label{GVrelation}
\sum_{g=0}^\infty \epsilon_-^{2g-2} F_g
= \log\left[ {\rm Tr} (-1)^F e^{-2\epsilon_- J_-^3} e^{-\beta H} \right]. 
\end{equation}
For toric Calabi-Yau 3-folds, the topological vertex constructed 
in \cite{topvertex} can be used to show this identify without 
relying on the heterotic dual or the $S \rightarrow \infty$ limit. 

Note that the right-hand side of this equation is the vacuum amplitude in the ${\cal N}=2$ theory
computed in the self-dual ($\epsilon_+=0$) Omega background. The left-hand side, however, is not
a vacuum amplitude, but it is a series of F-terms evaluated on the self-dual graviphoton background
before performing the superspace integral. If we were computing the vacuum amplitude in the left-hand side, 
we should have included the additional combinatorial factor of $(g!)^2$, among other possible contributions. 

\section{3. Supersymmetric configurations with field strengths in four dimensions}

When $\epsilon_+=0$, Nekrasov's partition function is the vacuum amplitude on the self-dual 
Omega background, and it is related to the F-terms for the self-dual graviphoton background. 
To find their generalizations for $\epsilon_+ \neq 0$, it is natural 
to look for configurations with both self-dual and anti-self-dual field strengths turned on. 
As we will see in this section, even the 
self-dual Omega background involves a gauge field other than the graviphoton when interpreted as a 
supergravity configuration. 
The case with $\epsilon_+ \neq 0$ will involve even more gauge fields. 

We assume massless hypermultiplets are all uncharged and their scalar
components are constant. With $n_V$ vector multiplets, the supersymmetry 
transformation laws for the gravitini $\psi_{\alpha\dot\alpha
 \beta}^i$ and 
gaugini $\lambda_\alpha^{A i}$
($i=1,2; A=1,..., n_V$) are \cite{de Wit:1984px,Andrianopoli:1996cm},
\begin{equation}\label{susylaw}
\begin{split}
&\delta \psi_{\alpha\dot\alpha\beta}^i = \nabla_{\alpha\dot\alpha} \zeta_\beta^i
 + F^G_{\alpha\beta} \zeta^i_{\dot\alpha}, \\
&\delta \psi_{\alpha\dot\alpha\dot\beta}^i = \nabla_{\alpha\dot\alpha} \zeta_{\dot \beta}^i
 + F^{\bar{G}}_{\dot\alpha\dot\beta} \zeta^i_{\alpha}, \\
&\delta \lambda_\alpha^{Ai} = \partial_{\alpha\dot \alpha} t^A \zeta^{\dot \alpha i}+
(F^A_{\alpha\beta} \delta^i_j + \epsilon_{\alpha\beta} P^{Ai}_j) \zeta^{\beta j}, \\
&\delta \lambda_{\dot \alpha}^{\bar Ai} = \partial_{\alpha\dot \alpha} t^{\bar A} \zeta^{\alpha i} +
(F^{\bar A}_{\dot \alpha\dot \beta} \delta^i_j + \epsilon_{\dot \alpha\dot \beta} P^{\bar Ai}_j) \zeta^{\dot \beta j},
\end{split}
\end{equation}
where $(\zeta_\alpha^i, \zeta_{\dot \alpha}^i)$  are supersymmetry transformation parameters, 
$F^G$ is the graviphoton field strength, $t^A$ is a vector multiplet scalar, $F^A$ is the corresponding
gauge field strength, and $\epsilon_{\alpha\beta}, \epsilon_{\dot\alpha\dot\beta}$ are antisymmetric bispinors
normalized as $\epsilon_{12}=\epsilon_{\dot 1 \dot 2}=1$. We also allowed FI parameters $P^A$
for the vector multiplets. 
Our task is to find configurations that are annihilated by some of these transformations. 

\subsection{3.1. Configurations without FI terms}

Let us start with configurations without FI terms. We assume that the graviphoton is self-dual
and vector multiplet fields are either self-dual or anti-self-dual.

\subsubsection{3.1.1 $F_{\alpha\beta}^G \neq 0$ and all other $=0$}

This is the background for the original topological string theory
studied in \cite{Bershadsky:1993cx,Antoniadis:1993ze}.
In this case, non-trivial parts of the supersymmetry transformation laws are
\begin{equation}
\begin{split}
&\delta \psi_{\alpha\dot\alpha\beta}^i = \partial_{\alpha\dot\alpha} \zeta_\beta^i
 + F^G_{\alpha\beta} \zeta^i_{\dot\alpha}, \\
&\delta \psi_{\alpha\dot\alpha\dot\beta}^i = \partial_{\alpha\dot\alpha} \zeta_{\dot \beta}^i. 
\end{split}
\end{equation}
If the self-dual graviphoton field strength $F_{\alpha\beta}^G$ is constant,
the right-hand sides of these equations can be set to zero by choosing,
\begin{equation}
\begin{split}
& \zeta_\alpha^i = \zeta_\alpha^{(0)i} - F_{\alpha\beta}^G x^{\beta \dot \beta} \zeta_{\dot \beta}^{(0)i}, \\
& \zeta_{\dot \alpha}^i = \zeta_{\dot \alpha}^{(0) i},
\end{split}
\end{equation}
where $(\zeta_\alpha^{(0)i}, \zeta_{\dot \alpha}^{(0)i})$ are constant spinors. 

Interestingly, this background preserves the same amount of supersymmetry as the flat Minkowski space 
does \cite{Ooguri:2003qp, Berkovits:2003kj}.  
In particular, the Hamiltonian is still expressed as an anti-commutator of the supercharges.
The supersymmetry algebra is modified as \cite{Berkovits:2003kj},
\begin{equation}\label{modifiedsusy}
\begin{split}
& \{ Q_\alpha^i, Q_{\dot \beta}^j \} = 2 \epsilon^{ij} P_{\alpha \dot \beta}, \\
& [ P_{\alpha\dot \alpha}, Q_{\dot \beta}^i ]
= 2 \epsilon_{\dot \alpha \dot \beta} F_{\alpha\beta}^G Q^{\beta i}, \\
& \{ Q_{\dot \alpha}^i ,Q_{\dot \beta}^j \} = 4 \epsilon_{\dot \alpha \dot \beta} \epsilon^{ij} F_{\alpha\beta}^G 
M^{\alpha\beta}, 
\end{split}
\end{equation}
where $P_{\alpha\dot\alpha}$ and $M^{\alpha\beta}$ are the momentum and the angular momentum generators.  

\subsubsection{3.1.2 $F_{\alpha\beta}^G, F_{\dot \alpha \dot \beta}^{\bar A} \neq 0$ and all other $=0$}

This is the configuration proposed in \cite{Antoniadis:2010iq} for a worldsheet description of
Nekrasov's partition function with $\epsilon_+ \neq 0$. The relevant supersymmetry transformation laws are,
\begin{equation}
\begin{split}
&\delta \psi_{\alpha\dot\alpha\beta}^i = \partial_{\alpha\dot\alpha} \zeta_\beta^i
 + F^G_{\alpha\beta} \zeta^i_{\dot\alpha}, \\
&\delta \psi_{\alpha\dot\alpha\dot\beta}^i = \partial_{\alpha\dot\alpha} \zeta_{\dot \beta}^i, \\
&\delta \lambda_{\dot \alpha}^{\bar A i} = F_{\dot \alpha \dot \beta}^{\bar A} \zeta^{\dot \beta i}. 
\end{split}
\end{equation}
Assuming $\det F^{\bar{A}}_{\dot \alpha \dot \beta}\neq 0$ for some gauge fields, the remaining supersymmetry 
is parametrized as $\zeta_\alpha^i = \zeta_\alpha^{(0) i}, \zeta_{\dot \alpha}^i = 0$. 
This configuration preserves all of the $Q_\alpha^i$ supercharges, but 
breaks the $Q_{\dot \alpha}^i$ supersymmetry. 

\subsubsection{3.1.3 $F_{\alpha\beta}^G, F_{\alpha \beta}^{A}, \partial t^A \neq 0$ and all other $=0$}

In this case, the relevant supersymmetry transformation laws are,
\begin{equation}\label{deformedsusy}
\begin{split}
&\delta \psi_{\alpha\dot\alpha\beta}^i = \partial_{\alpha\dot\alpha} \zeta_\beta^i
 + F^G_{\alpha\beta} \zeta^i_{\dot\alpha}, \\
&\delta \psi_{\alpha\dot\alpha\dot\beta}^i = \partial_{\alpha\dot\alpha} \zeta_{\dot \beta}^i, \\
&\delta \lambda_{\alpha}^{A i} = \partial_{\alpha \dot \alpha} t^A \zeta^{\dot \alpha i} + 
F_{\alpha \beta}^{A} \zeta^{\beta i}. 
\end{split}
\end{equation}
Assuming that $F_{\alpha\beta}^G$ is constant, we can
set $\delta \psi_{\alpha\dot\alpha \beta}^i = 0$ and $\delta \psi_{\alpha\dot\alpha\dot\beta}^i=0$
by parametrizing $\zeta^i_\alpha, \zeta^i_{\dot \alpha}$ as,
\begin{equation}\begin{split}
&\zeta_\alpha^i = - F_{\alpha\beta}^G x^{\beta \dot \beta} \zeta_{\dot \beta}^{(0)i}, \\
&\zeta_{\dot \alpha}^i = \zeta_{\dot \alpha}^{(0) i}.
\end{split}
\end{equation}
If $[F^G, F^A]_\alpha^\beta = 0$, we can also solve $\delta \lambda_\alpha^{Ai}=0$
by requiring that the vector multiplet moduli $t^A$ is position dependent as,
\begin{equation} \label{positiondependence}
t^A = t^A_0 + \frac{1}{2} (F^G x)^{\alpha\dot\alpha} (F^A x)_{\alpha \dot \alpha}. 
\end{equation}

This configuration, however, does not solve the supergravity equations of motion. 
The field strength $F^A$ is  defined as,
\begin{equation}
  F_A = \partial_A {\mathbb X}^I F_I, ~~(A=1,...,n_V),
\end{equation}
where ${\mathbb X}^I$ ($I=0,1,...,n_V$) are projective coordinates of the vector multiplet moduli space,
and $F_I$'s are the field strengths that satisfy the standard Bianchi identities, $d F_I=0$.
Therefore, if the moduli scalar fields $t^A$ are position dependent as in (\ref{positiondependence}),
$F^A = {\rm const.}$ is not compatible with the Bianchi identities. 

Fortunately, we know how to modify the field configuration to solve the equations of motion
for a case relevant to our problem. 
Recall that the Kaluza-Klein ansatz (\ref{consthetero}) with the gauge field
${\bar A}_\mu=0$ gives 
\begin{equation}\label{Kaluzatwo}
\begin{split}
ds^2 & = dx^2 + dz (d\bar z + 2A^\mu dx^\mu) \\
& = dx^2 + 2A_\mu dx^\mu dz + dz d\bar z. 
\end{split}
\end{equation}
From (\ref{GSTUidentification}), we see that the self-dual $F = dA$ corresponds to 
turning on $F_{\alpha\beta}^G = F_{\alpha\beta}^T=\frac{1}{2} F_{\alpha \beta}$. 

On the other hand, the Melvin-type geometry (\ref{Melvin}) has the metric,  
\begin{equation} \label{Melvintwo}
\begin{split}
ds^2 & = (dx^\mu + \Omega^\mu dz)^2 + dz d\bar z \\
& = dx^2 + 2 \Omega_\mu dx^\mu dz +  \Omega_\mu \Omega^\mu dz dz + dz d\bar{z}.
\end{split}
\end{equation}
Comparing the two metrics,
we see that the complex structure moduli field $t^T$ for the Melvin-type geometry,
appearing in the coefficient of $dz dz$ in the metric, has position dependence 
in four dimensions as
\begin{equation}
 t^T = t_0^T +  \frac{1}{2} \Omega^{\mu} \Omega_\mu \label{modulidep},
\end{equation}
where $t_0^T$ is the complex moduli of the original $T^2$. 
If we choose the gauge, 
\begin{equation} \label{Melvingauge}
  A_{\alpha\dot \beta}^G = A_{\alpha\dot \beta}^T 
  = \Omega_{\alpha\dot \beta} = F_{\alpha \beta} x^\beta_{~\dot \beta}, 
\end{equation}
the moduli field $t^T$ (\ref{modulidep}) shows 
the same position dependence as in (\ref{positiondependence}).

Therefore, to the leading nontrivial order, the Melvin-type background
agrees with the configuration of $F_{\alpha\beta}^G, F_{\alpha\beta}^T, \partial t^T \neq 0$
discussed in this section. The Melvin background with (\ref{Melvingauge}) is an exact
solution to the supergravity equations of motion since it is locally trivial. 
Moreover, the supersymmetry (\ref{deformedsusy}) is exactly the same as the one preserved
by the Melvin-type background \cite{Nekrasov:2003rj}. Thus, we conclude that the 
Melvin-type background gives a completion of the configuration discussed here, 
when the self-dual field strength is turned on the $T$ direction (the complex moduli of $T^2$ for
the heterotic string on $T^2 \times K3$).

\subsection{3.2. Configurations with FI terms}

The definition of Nekrasov's partition function (\ref{nekrasov}) involves twisting by the $SU(2)$ R symmetry. 
Type II superstring theory compactified on a Calabi-Yau manifold $X$ is R symmetric only if 
$X$ has a special isometry generated by the Reeb vector \cite{Martelli:2006yb}. In particular, $X$ must be non-compact. 
This is expected from the fact that there is no global symmetry in a quantum gravity theory with finite
Newton constant. 
Similarly, for the heterotic string on $T^2 \times K3$, the R symmetry requires $K3$ is non-compact and has 
an isometry generated by the Reeb vector, which 
rotates the covariantly constant spinor on $K3$. 
For example, an isolated ADE singularity (e.g. orbifold of $\mathbb{C}^2$) has such an isometry. 

The twisting by the R symmetry is closely related to turning on of the 
FI parameters $P^{Ai}_j$,
which is in the adjoint representation of the $SU(2)$ R symmetry ($i,j=1,2$). 
When the internal space is non-compact and allows the R symmetry in four dimensions, 
one can construct vertex operators on the string worldsheet which correspond to 
turning on the FI terms. In the context
of the heterotic string theory, these vertex operators will be discussed 
in detail in the next section.  
There, it will be shown why the R twist can be realized by turning on the FI parameters.

It is known that FI parameters are quantized in a supergravity theory with finite Newton constant. 
However, continuous FI parameters are possible if the internal space is non-compact and the Newton constant
in four dimensions is zero. 

In the perturbative heterotic string theory, only the $U(1)$ part of 
the R symmetry is manifest 
while the full $SU(2)$ R symmetry is an accidental symmetry in the supergravity limit. 
Although we will use the $SU(2)_R$ notation in this section, our supergravity analysis is applicable 
even if only a $U(1)_R$ subgroup is manifest since we only need to turn on the R twist (and therefore  FI terms) for a particular $U(1)_R$ direction.

Before describing backgrounds with FI terms, we would like to comment on analytic continuations of field 
configurations. In string theory, we often allow moduli fields $t^A$ and their complex conjugates $t^{\bar A}$ as
independent parameters. For example, in extracting the Gromov-Witten invariants from the topological string theory,
one takes the limit of $\bar t \rightarrow \infty$ for the K\"ahler moduli while keeping $t$ finite. 
This is allowed since there are separate vertex operators for $t^A$ and $t^{\bar A}$. Similarly, 
turning on self-dual field strength $F_{\alpha\beta}$ while keeping $F_{\dot \alpha \dot \beta} =0$ is allowed since there are separate vertex operators
for $F_{\alpha\beta}$ and $F_{\dot \alpha\dot \beta}$ on the
worldsheet. This is not the case for the FI terms. For them, there is only one vertex operator for each vector
multiplet and we cannot treat $P^A$ and $P^{\bar A}$ as independent. If we turn on $P^A$, we must turn on
$P^{\bar A}$ simultaneously. This will become important in some of the cases below. 

\subsubsection{3.2.1. $F_{\alpha\beta}^G, F_{\alpha \beta}^{A}, P^A \neq 0$ and all other $=0$}
 
Let us start with a warm-up exercise with only self-dual field strengths and FI terms turned on. 
The relevant supersymmetry transformation laws in this case are, 
\begin{equation}\label{susylaw}
\begin{split}
&\delta \psi_{\alpha\dot\alpha\beta}^i = \nabla_{\alpha\dot\alpha} \zeta_\beta^i
 + F^G_{\alpha\beta} \zeta^i_{\dot\alpha}, \\
&\delta \psi_{\alpha\dot\alpha\dot\beta}^i = \nabla_{\alpha\dot\alpha} \zeta_{\dot \beta}^i
 , \\
&\delta \lambda_\alpha^{Ai} = 
(F^A_{\alpha\beta} \delta^i_j + \epsilon_{\alpha\beta} P^{Ai}_j) \zeta^{\beta j}.
\end{split}
\end{equation}
When the FI terms are non-zero, we need to take into account the $SU(2)_R$ connection 
$V_{\mu j}^i = A_\mu^A P_{Aj}^i$ in the covariant derivative $\nabla_{\alpha\dot \alpha}$ in the first and second lines. 
In addition, we will need to turn on the spin connection in four dimensions. 

By definition, the FI terms $P^{Ai}_j$ are in the adjoint representation of the $SU(2)$ R symmetry
($i,j =1,2$; hermitian and traceless) and commute with each other, $[P^A, P^B] = 0$. 
On the other hand, $F_\alpha^{A\beta}$ is also in the adjoint representation of $SU(2)_- \in SO(4)$ for
the four-dimensional rotation.  If we turn on the FI terms so that
$\det P^A = \det F^A$, the eigenvalues of $P^A$ and $F^A$ match and 
the rank of the $4 \times 4$ matrix $(F_{\alpha\beta}^A \delta^i_j + \epsilon_{\alpha\beta} P^{Ai}_j)$
becomes $2$. Therefore, there are solutions to $(F^A + P^A)\zeta^{(0)} = 0$. 
Moreover, if we require the field strengths to commute $[F^A, F^B] = 0$, we have that $[F^A + P^A, F^B + P^B ] = 0$, 
and there are simultaneous solutions to $(F^A + P^A)\zeta^{(0)} = 0$ for all directions of the vector multiplets $A = 1, \cdots , n_V$. 

The integrability of $\nabla_{\alpha\dot \alpha} \zeta_{\beta}^i = 0$ requires that we tune the spin connection to cancel the effect
of the $SU(2)_R$ connection $V_{\mu j}^i = A_\mu^A P_{Aj}^i$. This turns on the self-dual component of the Riemann curvature as,
\begin{equation} 
 R_{\alpha\beta\gamma\delta} \sim F_{\alpha\beta}^A F_{A \gamma\delta}. 
\end{equation}

The traceless component of the Ricci tensor is zero, as expected from the Einstein equation since 
the energy-momentum tensor vanishes for the self-dual gauge fields.  
However, the scalar curvature is non-zero, 
\begin{equation}
R \sim \epsilon^{\alpha\gamma} \epsilon^{\beta\delta} R_{\alpha\beta\gamma\delta}
\sim F_{\alpha\beta}^A F_A^{\alpha\beta} = P^{Ai}_j P_{Ai}^j. 
\end{equation}
This is expected since the FI terms contribute to the cosmological constant. 
To understand why the FI parameters survive in the Einstein equation
when the Newton constant in four dimensions is zero,
it is convenient to restore mass dimensions of various parameters. 
The field strength $F^A$ and the FI parameter $P^A$ are of dimension 2,
while the $\epsilon_\pm$ parameters are of dimension 1. Thus, $\epsilon_\pm$ must be of the order of $P^A/M_{{\rm Planck}}$,
where $M_{{\rm Planck}}$ is the Planck mass in four dimensions. 
The Einstein equation should then set the Ricci curvature to be related to $\epsilon_\pm^2$
without an extra factor of $M_{{\rm Planck}}$, and the effect survives in the limit of $M_{{\rm Planck}} 
\rightarrow \infty$. This scaling is consistent with the supergravity 
transformation \eqref{susylaw} and the commutation relations \eqref{modifiedsusy}.

The configuration preserves the supersymmetry
\begin{equation}
\zeta_{\alpha}^i = \zeta_{\alpha}^{(0)i},
\ \zeta_{\dot{\alpha}}^i = 0,
\end{equation}
which gives rise to half of the $Q_{\alpha}^i$ charges.


%
%

\subsubsection{3.2.2. $F_{\alpha\beta}^G, F_{\alpha\beta}^A,  F_{\dot \alpha \dot \beta}^{\bar A}, P^A, \partial t^A \neq 0$ and all other $=0$}

This is the most general case we consider in this paper. We turn on almost all the fields, 
except for the anti-self-dual components of 
the graviphoton $F_{\dot \alpha \dot \beta}^G$ and the position dependence of the anti-holomorphic parts of the 
moduli fields $\partial \bar t^{\bar A}$. The relevant supersymmetry transformation laws are,
\begin{equation}\label{finalcase}
\begin{split}
&\delta \psi_{\alpha\dot\alpha\beta}^i = \nabla_{\alpha\dot\alpha} \zeta_\beta^i
 + F^G_{\alpha\beta} \zeta^i_{\dot\alpha}, \\
&\delta \psi_{\alpha\dot\alpha\dot\beta}^i = \nabla_{\alpha\dot\alpha} \zeta_{\dot \beta}^i, \\
&\delta \lambda_\alpha^{Ai} = \partial_{\alpha\dot \alpha} t^A \zeta^{\dot \alpha i}
+ (F^A_{\alpha\beta} + P^{Ai}_j) \zeta_\alpha^j, \\
&\delta \lambda_{\dot \alpha}^{\bar Ai} =
(F^{\bar A}_{\dot \alpha\dot \beta} \delta^i_j + \epsilon_{\dot \alpha\dot \beta} P^{\bar Ai}_j) \zeta^{\dot \beta j}.
\end{split}
\end{equation}
We assume that the field strengths commute, $[F^G, F^A] = 0, [F^A, F^B] = 0$, and $[F^{\bar A} F^{\bar B}] = 0$.

If we set ${\rm det} F^A = {\rm det} P^A$ and ${\rm det} F^{\bar A} = {\rm det} P^{\bar A}$ simultaneously,
we can find a configuration which preserves half of the $Q_{\alpha}^i$ supercharges and half of the $Q_{\dot \alpha}^i$ supercharges. 
To see this, choose constant spinors $\zeta_\alpha^{(0)i}$ and $\zeta_{\dot \alpha}^{(0)i}$ so that 
they are annihilated by $(F^A + P^A)$ and $(F^{\bar{A}} + P^{\bar{A}})$. 
Using
\begin{equation}\begin{split}
 \zeta_\alpha^i & = \zeta_\alpha^{(0)i} - F_{\alpha\beta}^G x^{\beta\dot\beta} \zeta^{(0) \dot\beta i}, \\
 \zeta_{\dot \alpha}^i & = \zeta_{\dot \alpha}^{(0) i}, 
 \end{split}
 \end{equation}
 for the supersymmetry transformation laws (\ref{finalcase}), we find that $\delta \lambda^{\bar{A}i}_{\dot \alpha}=0$ is
 automatic and $\delta \lambda^{Ai}_{\alpha}=0$ can be satisfied if we set 
 \begin{equation}
 t^A = t_0^A + \frac{1}{2}(F_\alpha^{G\beta} x_{\beta\dot \beta} F_\gamma^{A\alpha} x^{\gamma\dot \beta}
 - F^G_{\alpha\beta} x^{\beta\dot\beta} F_{\dot\beta \dot \gamma}^A x^{\dot \gamma \alpha}).
 \end{equation}
 The integrability of $\nabla_{\alpha\dot\alpha}$ requires
 \begin{equation}
  R_{\dot \alpha\dot\beta\dot \gamma\dot \delta} \sim F^A_{\dot \alpha \dot \beta} F_{A \dot \gamma \dot \delta},
  \end{equation}
  and that the scalar curvature is proportional to $P^A P_A$, which is the trace part of the Einstein 
  equation. In this case, the integrability further requires
  \begin{equation} 
  R_{\alpha\beta \dot \gamma \dot \delta} \sim F_{\alpha\beta}^A F_{A \dot \gamma \dot \delta}.
  \end{equation}
  The left-hand side is the traceless part of the Ricci curvature and the right-hand side is the energy-momentum tensor
for the gauge field, which is non-zero in this case since both self-dual and anti-self-dual components are turned on.
This is consistent with the Einstein equation. Note that the moduli scalar fields $t^A$ do not contribute to the 
energy-momentum tensor even though they are position dependent since we assume $\partial t^{\bar A} = 0$. 

As in the case we saw in section 3.1.3, the position dependence of $t^A$ means that constant $F^A$ does not satisfy the Bianchi
identities. Thus, we need modify the field configuration order by order in an expansion in powers of
the gauge field strengths so that the equations of motions are solved while preserving the same set of supercharges. 
In the next section, we will propose that the background appropriate for the refined topological string computation 
for the case with $\epsilon_+\neq 0$ is a configuration of this type. 

We can also relax the conditions, ${\rm det} F^A = {\rm det} P^A$ and ${\rm det} F^{\bar{A}} = {\rm det} P^{\bar{A}}$,
and find a configuration with less supersymmetry. We will show that the Omega background with ($\epsilon_+\neq 0$) 
is a configuration of this type.
 
 \subsection{3.3. Summary}
 
Let us summarize the supersymmetric configuration studied in this section. 
\begin{center}
\begin{tabular}{c|c|c|c}
 fields turned on   & $ \zeta_{\alpha}^{(0) i} $& $ \zeta_{\dot \alpha}^{(0) i}$ &   \\
 \hline
   $F_{\alpha \beta}^G$&     $ 1 $         &$1$ & TST ($\epsilon_+=0$)\\
  $F_{\alpha \beta}^G, F_{\dot \alpha \dot \beta}^{\bar{A}} $&     $1 $ &$ 0 $& discussed in \cite{Antoniadis:2010iq} \\  
  $F_{\alpha \beta}^G, F_{\alpha\beta}^A,  \partial t^A $&     $0 $ &$ 1 $& Omega ($\epsilon_+=0$)\\
$F_{\alpha \beta}^G, F_{\alpha\beta}^A, F_{\dot \alpha \dot \beta}^{\bar{A}}, P^A, \partial t^A $& $0 $ &$ 1/2 $& Omega ($\epsilon_+\neq 0$)\\
$F_{\alpha \beta}^G, F_{\alpha\beta}^A, F_{\dot \alpha \dot \beta}^{\bar{A}}, P^A, \partial t^A $& $1/2 $ &$ 1/2 $& TST
 ($\epsilon_+\neq 0$)\\
 \end{tabular}
\end{center}
The middle two columns count the number of supercharges preserved. 
On the right column, TST refers to the topological string theory and Omega is for the 
Omega background.  
The last two rows are the ones discussed in subsection 3.2.2. In the next section,
we will propose them as the Omega background and the topological 
string background for $\epsilon_+ \neq 0$. We did not list the configuration discussed in subsection 3.2.1
as it was meant to be a warm-up exercise.

\section{4. Omega Background and Refined Topological String Theory}

In this section, we will embed the backgrounds discussed in the previous section in the heterotic string theory on $T^2 \times K3$
and identify the Omega background with $\epsilon_+ \neq 0$. We will also propose the background for the corresponding topological
string computation. In the next section, we will discuss interpretations of these backgrounds in the context of type II superstring
theory on a Calabi-Yau manifold.

\subsection{4.1. Omega Background}

Identifying the supergravity description of the Omega background in $\mathbb{R}^4 \times T^2 \times K3$
is straightforward since the field configuration is locally trivial. The geometry is a flat bundle
of $\mathbb{R}^4 \times K3$ over $T^2$ so that, as we go along $T^2$, we perform the rotation of
$\mathbb{R}^4$ by the amount specified by $(\epsilon_-, \epsilon_+)$ and the isometry on $K3$
generated by the Reeb vector. 

When we reduce this 10-dimensional description of the Omega background down to 
$\mathbb{R}^4$, we find that it is a special case of the
configuration discussed in section 3.2.2. We first note that, to the linear order in $\epsilon_\pm$,
Melvin-type geometry, where $\mathbb{R}^4$ is fibered over $T^2$, 
is the same as the Kaluza-Klein geometry, where $T^2$ is fibered over
$\mathbb{R}^4$. We saw in (\ref{GSTUidentification}), 
such a Kaluza-Klein geometry for the heterotic string theory on $\mathbb{R}^4 \times T^2 \times K3$
requires turning on the $S$, $T$, and $U$ gauge fields
as well as the graviphoton field, as
\begin{equation}\label{generalomegadet}
\begin{split}
 & {\rm det} F_{\alpha\beta}^G = {\rm det} F_{\alpha\beta}^T = -\frac{1}{4} \epsilon_-^2, \\
  & {\rm det} F_{\dot \alpha\dot \beta}^{\bar{U}} = {\rm det} F_{\dot \alpha \dot \beta}^{\bar{S}} = -\frac{1}{4} \epsilon_+^2.
 \end{split}
 \end{equation}

Since both self-dual and anti-self-dual components of vector multiplet gauge fields are turned on, 
in order to preserve supersymmetry, we also need
to introduce FI terms. This corresponds to the R twist in Nekrasov's partition function
computation. Since the amount of the R twist is proportional to $\epsilon_+$, we must set $P^{\bar{U}}$ and
$P^{\bar{S}}$ as,
\begin{equation}
{\rm det} P^{\bar{U}} = {\rm det} P^{\bar{S}} = -\frac{1}{4}\epsilon_+^2,
\end{equation}
so that $(F^{\bar{U}}+P^{\bar{U}})$ and $(F^{\bar{S}}+P^{\bar{S}})$ can annihilate some spinors.

As we saw in (\ref{GSTU}), the self-dual component $F_-^T$ and the anti-self-dual component $F_+^{\bar{U}}$
for the $T$ and $U$ fields come from the same gauge field. Therefore,
their FI terms are related as $P^{\bar{U}} = P^T$ and we must have 
${\rm det} P^T = -\frac{1}{4}\epsilon_+^2$. 
On the other hand, $F_{\alpha\beta}^T$ is set as
${\rm det} F^T= -\frac{1}{4} \epsilon_-^2$ by (\ref{generalomegadet}).
Therefore $(F^T + P^T)$ does not have a zero mode unless $\epsilon_+ = \pm \epsilon_-$. 

This configuration preserves half of the 
$Q_{\dot\alpha}^i$ supercharges but breaks all of
$Q_\alpha^i$. The constant field strengths of $F^T, F^{\bar{U}}$ and $F^{\bar{S}}$ do not
satisfy the Bianchi identities because of the position dependence of the moduli scalar field. 
We can modify the configuration order by order in $\epsilon_\pm$. The non-self-dual Melvin-type 
background with the R-twist, namely the non-self-dual Omega background, gives a completion of
this iterative procedure 
since it is an exact solution to the equations of motion and preserves the same set of supercharges.
 
 It is interesting to note that,
 when $\epsilon_+=\pm \epsilon_-$, we have zero modes for $(F^T + P^T)$ and half of the $Q_\alpha^i$ supercharges are restored.
 In fact, in this case, the Omega background twists only a 2-dimensional subspace in $\mathbb{R}^4$ keeping the transverse $\mathbb{R}^2$ intact. 
 This is known as the Nekrasov-Shatashvili limit \cite{Nekrasov:2009rc}.

\subsection{4.2. R twist and FI terms}

To describe the Omega background on the string worldsheet, we need to understand the relation between the R twist and 
the FI terms better. 
To construct the vertex operators for the FI terms, we assume that the $K3$ surface has a special isometry that generates the 
R symmetry. This means, in particular, that $K3$ is non-compact. In this case, there is a corresponding current $J_I^3$
on the left-moving (supersymmetric) sector of the worldsheet theory. However, it does not preserve the worldsheet
BRST symmetry by itself. To construct a physical vertex operator, we must combine it with one of the $SU(2)$ current $J_{N=4}^3$ in the 
${\cal N}=4$ superconformal algebra in the left-mover. The combination $(J_I^3 + J_{N=4}^3)$ is BRST invariant. 

If there is a current ${\cal J}$ with conformal weight $(0,1)$ in the right-moving (bosonic) sector, there is a
corresponding vector multiplet in four dimensions. For example, the vertex operator for the gauge field is 
$A_\mu \partial X^\mu {\cal J}$. The FI term for this multiplet is constructed using the BRST invariant R symmetry generator
as $(J_I^3 + J_{N=4}^3) {\cal J}$. In fact, one can verify that it is in the same supermultiplet as
the moduli scalar field for ${\cal J}$ by commuting these vertex operators with the target space supercharges as,
\begin{equation}
\begin{split}
\{Q_{\alpha}^1, [Q_\beta^2 ,(J_I^3 + J_{N=4}^3) {\cal J} ] \} = \epsilon_{\alpha \beta} e^{\varphi} \psi_Z {\cal J}.
\end{split} \label{commu}
\end{equation}
The right-hand side is the picture one vertex operator for the moduli scalar field, where $e^{\varphi}$ is the bosonized superghost.

The vertex operators for the $T$ and $U$ gauge fields are,
\begin{equation}
\begin{split}
& (A^T_\mu)_- \partial X^\mu \bar \partial Z, ~~ (A^{\bar{T}}_\mu)_+ \partial X^\mu \bar \partial \bar Z, \\
& (A^U_\mu)_- \partial X^\mu \bar \partial \bar Z, ~~ (A^{\bar{U}}_\mu)_+ \partial X^\mu \bar \partial Z. 
\end{split}
\end{equation}
This means that $(A_\mu^T)_-$ and $(A^{\bar{U}}_\mu)_+$ have the same FI term, 
$P^T = P^{\bar{U}}$, which couples to the vertex operator
$(J_I^3 + J_{N=4}^3) \bar \partial Z$. Similarly, $(A_\mu^{\bar{T}})_+$ and 
$(A^U_\mu)_-$ have the same FI term, $P^{\bar{T}} = P^U$, 
with the vertex operator $(J_I^3 + J_{N=4}^3) \bar \partial \bar Z$. 

The vertex operators for $P^T$ and $P^{\bar{T}}$ constructed in this way
can be identified with the vertex operators for the $dz$ and $d\bar z$ components of
the Kaluza-Klein gauge field \eqref{twist} associated to the isometry
along the Reeb vector on $K3$. 
In this way, we understand that turning on the FI parameters 
is related to the R twist. This is analogous to the relation between the twisting parameter 
in the Melvin-type geometry and the introduction of the constant field strength to the 
four-dimensional Kaluza-Klein gauge field.


%

\subsection{4.3. Proposal for Refined Topological String Theory}

As we reviewed in section 2.2, when $\epsilon_+=0$, Nekrasov's partition function has the asymptotic expansion
which gives the all genus topological string partition function. The former is the partition function for the
Omega background, where $F_{\alpha\beta}^G = F_{\alpha\beta}^T$ are turned on, whereas the latter computes scattering amplitudes
for self-dual graviphotons, with 2 self-dual gravitons. In 4.1, we discussed the generalization
of the Omega background to the non-self-dual case. It is natural to ask what the corresponding generalization
of the topological string theory is. 
Following the standard terminology, we call it the refined topological string theory.  
Assuming that this string theory still computes some scattering amplitudes, we will try to identify what they are.
There are several constraints we need to take into account in order to make the identification. 

\medskip
\noindent
[1]
We require that $\epsilon_-$ and $\epsilon_+$ are self-dual and anti-self dual field strengths for
the Kaluza-Klein gauge field $g_{\mu z}$. Using the field identification (\ref{GSTU}), we find
\begin{equation}
\begin{split}
& \epsilon_- = (\partial_{[\mu,} g_{\nu] z})_- = F_-^G + F_-^T ,\\
& \epsilon_+ = (\partial_{[\mu,} g_{\nu] z})_+ = F_+^{\bar{U}} + F_+^{\bar{S}} .\\
\end{split}
\end{equation}
To minimize the set of assumptions, we require that no field strength other than 
$F_-^G, F_-^T, F_+^{\bar{U}}$ and $F_+^{\bar{S}}$ is turned on.
We also assume that only the FI terms for the $T$ and $U$ fields are 
turned on since they are the ones directly related to the R twist, as we saw in 4.2. 

\medskip
\noindent
[2]
We will find that $F_-^T$ has to be non-zero. Thus, both self-dual and anti-self-dual components
of vector multiplet gauge fields are turned on. 
We require that the relevant configuration preserves the maximally allowed 
amount of supersymmetry under the circumstance. In this case, it means half of the $Q_\alpha^i$ supersymmetry and
half of the $Q_{\dot\alpha}^i$ supersymmetry. 

We note that the general Omega background with $\epsilon_+\neq 0$ preserves half of the supersymmetry
of the self-dual Omega background with $\epsilon_+ =0$. Our proposal for the refined
topological string background preserves half of the supersymmetry of the self-dual graviphoton background
for the ordinary topological string theory. 

\medskip
\noindent
[3] The FI term for $F_-^T$ is the same as the FI term for $F_+^{\bar{U}}$. By the constraint [2], we must have
${\rm det} F_-^T = {\rm det} F_+^{\bar{U}}$. 

\medskip
\noindent
[4] Since $P^{\bar{S}} = 0$ by the assumption [1], we need to turn off $F_+^{\bar{S}}$. 

\medskip
Assuming these constraints, there is a unique choice:
\begin{equation}
\begin{split}  \label{background}
  F_-^G &= \epsilon_- - \epsilon_+, \\
  F_-^T &= F_+^{\bar{U}} = \epsilon_+, \\
 P^T  &=  P^{\bar{U}} = \epsilon_+ .
\end{split} 
\end{equation}
We note that, in the self-dual limit $\epsilon_+ = 0$,
this configuration reduces to the self-dual graviphoton background
for the original topological string theory.

We propose that the refined topological string theory computes
scattering amplitudes of the linear superpositions of various gauge fields
specified by (\ref{background}). More explicitly, 
let $A_{n_G, n_T, n_{\bar{U}}, n_P}^{\bar{n}_G, \bar{n}_T, \bar{n}_{\bar{U}}}$ be the scattering amplitude of 2 self-dual
 gravitons, $n_{G}+ \bar{n}_{G} -2$ self-dual graviphotons $F_{\alpha\beta}^G$, $n_T + \bar{n}_T$ self-dual vector fields 
 for $F_{\alpha\beta}^T$, $ n_{\bar{U}} + \bar{n}_{\bar{U}}$ anti-self-dual vector fields for $F_{\dot\alpha\dot\beta}^{\bar{U}}$, 
 and $n_P$ FI terms for $P^T =P^{\bar{U}}$, all at zero-momentum. Here the numbers with and without bar are 
 defined by the two different polarizations used in \cite{Bershadsky:1993cx,Antoniadis:1993ze} for the self-dual 
 case and \cite{Morales:1996bp,Antoniadis:2010iq} for the anti-self-dual case. (For a precise specification of
 the polarizations, see the expression in \eqref{actionn}). 
 In the unrefined case  \eqref{hetoneloop}, the worldsheet charge conservation dictates $n_G = \bar{n}_G$, whereas more generic 
 scattering amplitudes are non-zero here.
The generating function for scattering amplitudes for (\ref{background}) is then given by
\begin{equation}
\begin{split}  \label{generate}
&F(\epsilon_-,\epsilon_+) =    \sum_{g=0}^\infty g_s^{2g-2} F_g(b) \cr 
& = \underset{\bar{n}_{\bar{U}} , n_P = 0 }{\underset{ \bar{n}_T, n_{\bar{U}}, }{\sum_{n_G, \bar{n}_G, n_T,}^\infty}} \frac{(\epsilon_--\epsilon_+)^{n_G + \bar{n}_G} \epsilon_+^{n_T + \bar{n}_T + n_{\bar{U}} + \bar{n}_{\bar{U}} + n_P}}{(\epsilon_-^2 -\epsilon_+^2)(n_G ! \bar{n}_G! n_T ! \bar{n}_T! n_{\bar{U}} ! \bar{n}_{\bar{U}} ! n_P !)} A_{n_G, n_T, n_{\bar{U}}, n_P}^{\bar{n}_G, \bar{n}_T, \bar{n}_{\bar{U}}} \cr
& = \frac{1}{\epsilon_-^2 - \epsilon_+^2} \int d^2 \tau \langle e^{-\Delta S_{\mathrm{eff}}}  \rangle_{R^4 \times T^2 \times K3},
\end{split}
\end{equation}  
where we recall $\epsilon_+ = \frac{i}{2}g_s (b+b^{-1})$, $\epsilon_- = \frac{i}{2} g_s (b-b^{-1})$. The deformation of the worldsheet action $\Delta S_{\mathrm{eff}}$ is given by
\begin{equation} \label{actionn}
\begin{split}
\Delta S_{\mathrm{eff}} = &(\epsilon_- - \epsilon_+) \int \left[ (X^1 \partial Z - \chi^1 \psi_Z)\bar{\partial} X^2  \right. \\
&~~~~~~~~~~~~~~+ \left. (\bar{X}^1 \partial Z - \bar{\chi}^1 \psi_Z ) \bar{\partial} \bar{X}^2 \right) \\
&+ \epsilon_+ \int \left(X^1 \partial X^2 - \chi^1 \chi^2 + \bar{X}^1 \partial \bar{X}^2 - \bar{\chi}^1 \bar{\chi}^2 \right) \bar{\partial} Z \\
&+\epsilon_+ \int \left(\bar{X}^1 \partial X^2 - \bar{\chi}^1 \chi^2 + {X}^1 \partial \bar{X}^2 - {\chi}^1 \bar{\chi}^2 \right) \bar{\partial} Z \\
&+ \epsilon_+\int (J^3_I + J^3_{N=4}) \bar{\partial} Z + \mathcal{O}(\epsilon^2),
\end{split}
\end{equation}
which corresponds to the configuration \eqref{background}.
This is a generalization of \eqref{hetoneloop} for the non-self-dual background. The particular polarization and the combinatorial factor $(n_G ! \bar{n}_G! n_T ! \bar{n}_T! n_{\bar{U}} ! \bar{n}_{\bar{U}} ! n_P !)^{-1}$
in $(\ref{generate})$ are needed in order for the effective worldsheet action \eqref{actionn} to preserve 
the supersymmetry discussed in section 3.2.2.

As discussed in section 3.2.2, we need to turn on $\partial t^T$ at $\mathcal{O}(\epsilon^2)$.
 In addition, the four-dimensional space should be curved appropriately to preserve the supersymmetry. 
The requirement for the position dependence of $t^T$ can also be traced to the worldsheet operator product expansion (OPE) and renormalization. 
The OPE of the vertex operators $F_-^G$ and $F_-^T$ generates the complex structure moduli 
$\partial Z \bar{\partial} Z$ for the moduli field $t^T$. 
To preserve the worldsheet conformal invariance, we have to add the shift of $t^T$ to cancel this effect, and
the resulting position dependence of $t^T$ is precisely as required in section 3.2.2. 
Once this is taken into account, no more singularities are generated since there are no other 
operators that can have nontrivial OPE with $\bar{\partial}Z$.


\subsection{4.4. Test of the Proposal}
Although we have not been able to solve the worldsheet theory for the background proposed in the above, 
we can discuss aspects of the proposed string amplitudes. In particular, we present a modest 
test of this proposal in the zero-slope limit of the heterotic string, 
where the one-loop scattering amplitude and the corresponding generating function \eqref{generate} 
can be computed. This is the limit where Nekrasov's partition function is originally defined and computed from the field theory.

The contribution from an elementary hypermultiplet is,
\begin{equation}
\int_0^\infty \frac{dt}{t} \frac{ e^{-\mu t}}
{{\sinh} (\epsilon_- + \epsilon_+) t \ {\sinh}(\epsilon_- - \epsilon_+) t }.
\end{equation}
In this case, there is no effect of the R twist (i.e., the FI term), and 
the computation is essentially the same as the one in \cite{Antoniadis:2010iq,Nakayama:2010ff,Huang:2010kf}.
On the other hand, the R twist is relevant for an elementary vector multiplet, and its one-loop contribution is, 
\begin{equation}
\int_0^\infty \frac{dt}{t} \frac{ - 2 {\rm cosh}(2\epsilon_+ t) e^{-\mu t}}
{{\sinh} (\epsilon_- + \epsilon_+) t \ {\sinh}(\epsilon_- - \epsilon_+) t }.
\end{equation}
The extra $-2\cosh(2\epsilon_+ t)$ is precisely due to the FI term that gives the worldsheet R symmetry twist in the left-moving ground states.
These reproduce the contributions of the hyper and vector multiplets to Nekrasov's partition function \eqref{nekrasov}, and agree with the general formula,
\begin{equation}
\log Z(\epsilon_+, \epsilon_-) = \int \frac{dt}{t} \frac{\mathrm{tr} (-1)^{2J_-^3 + 2J_+^3}  e^{-4t \epsilon_- J^3_- - 4t \epsilon_+ J^3_+}}{{\sinh} (\epsilon_- + \epsilon_+) t \ {\sinh}(\epsilon_- - \epsilon_+) t } e^{-\mu t} 
\end{equation}
proposed in \cite{Hollowood:2003cv,Iqbal:2007ii}.

Our proposal also reproduces some of generic features of Nekrasov's partition function. FI term $P^T$ contains 
$(J_{N=4}^3+ J_I^3)$, and it generates twisting of the worldsheet R symmetry.
This leads to the twisting of the target space R symmetry via the conformal field theory for $K3$.  
Secondly, in the field theory limit, the amplitude has the manifest reflection symmetry 
$\epsilon_\pm \to -\epsilon_\pm$ because the $U(1)$ R symmetry is enhanced to the $SU(2)$ R symmetry.  
This reflection symmetry is an important feature of Nekrasov's partition function (see e.g. \cite{Iqbal:2007ii,Krefl:2010jb}).

\section{5. Type II interpretation}

In the last section, we proposed the worldsheet description of the generalized Omega background in the heterotic string theory
on $T^2 \times K3$. We performed a modest test of our proposal in the zero-slope limit and found that the one-loop heterotic
string computation reproduces Nekrasov's partition function in this limit. One way to test this proposal fully would be to compute 
the heterotic string one-loop amplitude exactly. Alternatively, one can perform the heat kernel expansion of higher-spin 
BPS multiplets and evaluate their contributions to the $R_-^2$ correction.
This may be tedious but is straightforward in principle. Work in this direction is in progress. 

By using the heterotic/type IIA string duality, we can transform the heterotic string background discussed in the last section 
into a type II superstring background. The Calabi-Yau three-fold in the dual type II background is non-compact, and the $U(1)$ R
symmetry necessary for the FI terms is geometrically realized by the Reeb vector. 
The corresponding vertex operators and their geometric interpretation have been proposed in \cite{Lawrence:2004zk,Aganagic:2007zm}. 

In the type II superstring theory, there is a separate set of supersymmetry for the left mover and right mover. 
Let us denoted them as $Q^1_{\alpha}$ and ${Q}^2_{\alpha}$ respectively. The vertex operator $V_{D^A}$ for the FI term
for the vector multiplet $A$ is characterized by,
\begin{align}
\{ {Q}^1_{\alpha}, [Q^2_\beta, V_{D^A}] \} = \epsilon_{\alpha\beta} V_{t^A}^{(-1,-1)} ,
\end{align}
where $V_{t^A}^{(-1,-1)}$ is the $(-1,-1)$ picture vertex operator for the corresponding moduli field $t^A$.
(Compare this with \eqref{commu} for the heterotic string theory.)
The FI vertex operator $V_{D_A}$ is in the RR sector, and its precise form in the type IIB string can be found in \cite{Lawrence:2004zk}. 
It is essentially the same vertex operator for the field strength $F^{A}_{\alpha\beta}$ for the gauge field except that the 4-dimensional 
part has antisymmetric rather than symmetric spin indices.

In the type II description, the worldsheet topological twist is generated by insertions of the self-dual 
graviphoton vertex operators \cite{Bershadsky:1993cx,Antoniadis:1993ze}. 
The anti-self-dual field strength $F^A_{\dot\alpha \dot \beta}$ required for $\epsilon_+\neq 0$ generates insertions of 
the (anti-chiral, anti-chiral) vertex operator $\Psi_A$ and 
the (chiral, chiral) vertex operator $\oint dz \rho(z) \oint d\bar z \bar\rho(\bar z) \Psi_A$,
where $\rho$ is the unique left-moving chiral operator of charge 3 which induces the spectral flow \cite{Antoniadis:2010iq}.
In addition, the FI term is represented by $(\oint dz \rho(z) - \oint d\bar{z} \rho(\bar{z})) \Psi_A$ 
if we use the RR vertex operator $V_{D^A}$ discussed in the above. Since the target space 
supersymmetry calls for simultaneous insertions of these three type of vertex operators, we
expect corresponding enhancement of symmetry in the worldsheet theory.

\section*{Acknowledgments}

We thank Nathan Berkovits, Nikita Nekrasov, Jaewon Song, 
and Cumrun Vafa for discussion.

This work is supported in part 
by U.S. Department of Energy grant DE-FG03-92-ER40701 and by the 
World Premier International Research Center Initiative of MEXT of
Japan. H.O. is also supported in part by Grant-in-Aid for Scientific 
Research C-20540256 and C-23540285 of Japan Society for the Promotion of 
Science.  



\begin{thebibliography}{99}
\bibitem{Bershadsky:1993cx}
  M.~Bershadsky, S.~Cecotti, H.~Ooguri, C.~Vafa,
  ``Kodaira-Spencer theory of gravity and exact results for quantum string amplitudes,''
  Commun.\ Math.\ Phys.\  {\bf 165}, 311-428 (1994). [hep-th/9309140].
\bibitem{Antoniadis:1993ze}
  I.~Antoniadis, E.~Gava, K.~S.~Narain, T.~R.~Taylor,
  ``Topological amplitudes in string theory,''
  Nucl.\ Phys.\  {\bf B413}, 162-184 (1994).  [hep-th/9307158].
\bibitem{Gopakumar:1998ii}
  R.~Gopakumar, C.~Vafa,
  ``M theory and topological strings. 1.,''
  [hep-th/9809187].
  R.~Gopakumar, C.~Vafa,
  ``M theory and topological strings. 2.,''
  [hep-th/9812127].

\bibitem{Ooguri:2004zv}
  H.~Ooguri, A.~Strominger, C.~Vafa,
  ``Black hole attractors and the topological string,''
  Phys.\ Rev.\  {\bf D70}, 106007 (2004).
  [hep-th/0405146].
\bibitem{Nekrasov:2002qd}
  N.~A.~Nekrasov,
  ``Seiberg-Witten prepotential from instanton counting,''
  Adv.\ Theor.\ Math.\ Phys.\  {\bf 7}, 831-864 (2004).
  [hep-th/0206161].

\bibitem{Losev:2003py}
  A.~S.~Losev, A.~Marshakov, N.~A.~Nekrasov,
  ``Small instantons, little strings and free fermions,''
  In *Shifman, M. (ed.) et al.: From fields to strings, vol. 1* 581-621.
  [hep-th/0302191].


\bibitem{Nekrasov:2003rj}
  N.~Nekrasov, A.~Okounkov,
  ``Seiberg-Witten theory and random partitions,''
  [hep-th/0306238].




\bibitem{Awata:2005fa}
  H.~Awata and H.~Kanno,
  ``Instanton counting, Macdonald functions and the moduli space of D-branes,''
  JHEP {\bf 0505}, 039 (2005)
  [arXiv:hep-th/0502061].


\bibitem{Iqbal:2007ii}
  A.~Iqbal, C.~Kozcaz, C.~Vafa,
  ``The Refined topological vertex,''
  JHEP {\bf 0910}, 069 (2009).
  [hep-th/0701156].
  

\bibitem{Dijkgraaf:2009pc}
  R.~Dijkgraaf, C.~Vafa,
  ``Toda Theories, Matrix Models, Topological Strings, and N=2 Gauge Systems,'
  [arXiv:0909.2453 [hep-th]].

\bibitem{Antoniadis:2010iq}
  I.~Antoniadis, S.~Hohenegger, K.~S.~Narain, T.~R.~Taylor,
  ``Deformed Topological Partition Function and Nekrasov Backgrounds,''
  Nucl.\ Phys.\  {\bf B838}, 253-265 (2010).
  [arXiv:1003.2832 [hep-th]].

\bibitem{Morales:1996bp}
  J.~F.~Morales, M.~Serone,
  ``Higher derivative F terms in N=2 strings,''
  Nucl.\ Phys.\  {\bf B481}, 389-402 (1996).
  [hep-th/9607193].

\bibitem{deWit:2010za}
  B.~de Wit, S.~Katmadas, M.~van Zalk,
  ``New supersymmetric higher-derivative couplings: Full N=2 superspace does not count!,''
  JHEP {\bf 1101}, 007 (2011).
  [arXiv:1010.2150 [hep-th]].
\bibitem{Ito:2010vx}
  K.~Ito, H.~Nakajima, T.~Saka, S.~Sasaki, H.~Nakajima, T.~Saka, S.~Sasaki,
  JHEP {\bf 1011}, 093 (2010).
  [arXiv:1009.1212 [hep-th]].


\bibitem{Antoniadis:1995zn}
  I.~Antoniadis, E.~Gava, K.~S.~Narain, T.~R.~Taylor,
  ``N=2 type II heterotic duality and higher derivative F terms,''
  Nucl.\ Phys.\  {\bf B455}, 109-130 (1995).
  [hep-th/9507115].

\bibitem{topvertex}
  M.~Aganagic, A.~Klemm, M.~Marino and C.~Vafa,
  ``The Topological vertex,''
  Commun.\ Math.\ Phys.\  {\bf 254}, 425 (2005)
  [arXiv:hep-th/0305132].

\bibitem{de Wit:1984px}
  B.~de Wit, P.~G.~Lauwers, A.~Van Proeyen,
  ``Lagrangians of N=2 Supergravity - Matter Systems,''
  Nucl.\ Phys.\  {\bf B255}, 569 (1985).

\bibitem{Andrianopoli:1996cm}
  L.~Andrianopoli, M.~Bertolini, A.~Ceresole, R.~D'Auria, S.~Ferrara, P.~Fre, T.~Magri,
  ``N=2 supergravity and N=2 superYang-Mills theory on general scalar manifolds: Symplectic covariance, gaugings and the momentum map,''
  J.\ Geom.\ Phys.\  {\bf 23}, 111-189 (1997).
  [arXiv:hep-th/9605032 [hep-th]].

\bibitem{Ooguri:2003qp}
  H.~Ooguri, C.~Vafa,
  ``The C deformation of Gluino and nonplanar diagrams,''
  Adv.\ Theor.\ Math.\ Phys.\  {\bf 7}, 53-85 (2003).
  [hep-th/0302109].

\bibitem{Berkovits:2003kj}
  N.~Berkovits, N.~Seiberg,
  ``Superstrings in graviphoton background and N=1/2 + 3/2 supersymmetry,''
  JHEP {\bf 0307}, 010 (2003).
  [hep-th/0306226].

\bibitem{Martelli:2006yb}
  D.~Martelli, J.~Sparks, S.-T.~Yau,
  ``Sasaki-Einstein manifolds and volume minimisation,''
  Commun.\ Math.\ Phys.\  {\bf 280}, 611-673 (2008).
  [hep-th/0603021].


\bibitem{Nekrasov:2009rc}
  N.~A.~Nekrasov, S.~L.~Shatashvili,
  ``Quantization of Integrable Systems and Four Dimensional Gauge Theories,''
  [arXiv:0908.4052 [hep-th]].

\bibitem{Nakayama:2010ff}
  Y.~Nakayama,
  ``Refined Cigar and Omega-deformed Conifold,''
  JHEP {\bf 1007}, 054 (2010).
  [arXiv:1004.2986 [hep-th]].

\bibitem{Huang:2010kf}
  M.-X.~Huang, A.~Klemm,
  ``Direct integration for general $\Omega$ backgrounds,''
  [arXiv:1009.1126 [hep-th]].

\bibitem{Hollowood:2003cv}
  T.~J.~Hollowood, A.~Iqbal, C.~Vafa,
  JHEP {\bf 0803}, 069 (2008).
  [hep-th/0310272].

\bibitem{Krefl:2010jb}
  D.~Krefl, J.~Walcher,
  ``Shift versus Extension in Refined Partition Functions,''
  [arXiv:1010.2635 [hep-th]].


\bibitem{Lawrence:2004zk}
  A.~Lawrence, J.~McGreevy,
  ``Local string models of soft supersymmetry breaking,''
  JHEP {\bf 0406}, 007 (2004).
  [hep-th/0401034].

\bibitem{Aganagic:2007zm}
  M.~Aganagic, C.~Beem,
  ``Geometric transitions and D-term SUSY breaking,''
  Nucl.\ Phys.\  {\bf B796}, 44-65 (2008).
  [arXiv:0711.0385 [hep-th]].


\end{thebibliography}
\end{document}